\documentclass[aps,prl,twocolumn,superscriptaddress,showpacs,showkeys]{revtex4-1}
\usepackage{graphicx}
\usepackage[abs]{overpic}
\usepackage{subfigure}
\newcommand{\bra}[1] {\langle #1 |}
\newcommand{\ket}[1] {| #1 \rangle}

\begin{document}

\title{Electronic properties and metrology of the diamond NV$^-$ center under pressure}
\author{Marcus W. Doherty}
\email{marcus.doherty@anu.edu.au}
\affiliation{Laser Physics Centre, Research School of Physics and Engineering, Australian National University, Australian Capital Territory 0200, Australia.}
\author{Viktor V. Struzhkin}
\affiliation{Geophysical Laboratory, Carnegie Institution of Washington, Washington DC 20015, USA.}
\author{David A. Simpson}
\affiliation{Centre for Quantum Computation and Communication Technology, School of Physics, University of Melbourne, Victoria 3010, Australia.}
\author{Liam P. McGuinness}
\affiliation{Centre for Quantum Computation and Communication Technology, School of Physics, University of Melbourne, Victoria 3010, Australia.}
\affiliation{Institute for Quantum Optics, University of Ulm, D-89081, Ulm, Germany}
\author{Yufei Meng}
\affiliation{Geophysical Laboratory, Carnegie Institution of Washington, Washington DC 20015, USA.}
\author{Alastair Stacey}
\affiliation{School of Physics and Melbourne Materials Institute, University of Melbourne, Victoria 3010, Australia.}
\author{Timothy J. Karle}
\affiliation{School of Physics and Melbourne Materials Institute, University of Melbourne, Victoria 3010, Australia.}
\author{Russell J. Hemley}
\affiliation{Geophysical Laboratory, Carnegie Institution of Washington, Washington DC 20015, USA.}
\author{Neil B. Manson}
\affiliation{Laser Physics Centre, Research School of Physics and Engineering, Australian National University, Australian Capital Territory 0200, Australia.}
\author{Lloyd C.L. Hollenberg}
\affiliation{Centre for Quantum Computation and Communication Technology, School of Physics, University of Melbourne, Victoria 3010, Australia.}
\author{Steven Prawer}
\affiliation{School of Physics and Melbourne Materials Institute, University of Melbourne, Victoria 3010, Australia.}

\date{\today}

\begin{abstract}
The negatively charged nitrogen-vacancy (NV$^-$) center in diamond has realized new frontiers in quantum technology.  Here, the center's optical and spin resonances are observed under hydrostatic pressures up to 60 GPa. Our observations motivate powerful new techniques to measure pressure and image high pressure magnetic and electric phenomena. Additionally, analysis of our observations provides clear insight into the pressure variation of the center's electronic orbitals.
\end{abstract}

%Localized modes, 63.20.Pw
%Phonon-defect interactions, 63.20.kp
%78.40.Ha = absorption spectra, color centers, nonmetallic inorganic, insulators
%76.70.hb = "ODMR" (replace?)
%61.72.jn = color centers, crystal defects (replace?)

\maketitle

The negatively charged nitrogen-vacancy (NV$^-$) center is a remarkable point defect in diamond and is the system of choice for a rapidly expanding domain of quantum technology demonstrations, including information processing \cite{wrachtrup06,maurer12}, communications \cite{beveratos02,alleaume04} and metrology \cite{taylor08,balasubramanian08,maze08,acosta10,dolde11,cole09,hall09,maclaurin12,ledbetter12,ajoy12}. The NV$^-$ center is also particularly rich in physics, however somewhat surprisingly, many fundamental aspects of the center remain unresolved (see Ref. \onlinecite{review} for an extensive review).  Indeed, most of center's applications exploit its unique combination of long-lived ground state spin coherence \cite{balasubramanian09,pham12} and optical spin-polarization/ readout (OSPR) that persist to ambient temperatures and beyond \cite{toyli12}. Attainment of a complete understanding of the latter is particularly important to the optimization of the NV$^-$ center's applications as well as to the identification of other useful defects.

Recently, the effects of temperature on the center's ground state spin and OSPR mechanism have been investigated, which enabled the influence of temperature on existing NV$^-$ metrology applications to be characterized and new thermometry applications to be proposed \cite{toyli12,acostashift1,acostashift2,chenshift}. There has been no commensurate investigation of the effects of pressure. Pressure and uniaxial stress enable the lattice and electronic orbitals of the NV$^-$ center to be manipulated, thus providing probes into the center's intriguing coupling of orbital and spin dynamics.  Such investigations will complete the characterization of NV$^-$ metrology under extreme conditions and is of particular interest to fields such as high pressure superconductivity, magnetic and electric phase transitions \cite{struz02,lin05,takahashi08,xchen10,sun12,kornev05,ahart10}. Here, we present the first observations of the behavior of the ground state spin of the NV$^-$ center under hydrostatic pressure. Our observations provide new insight into the physics of the center, motivate the center as a pressure sensor and enable an analysis of the center's other metrology applications under extreme conditions.

\begin{figure}[hbtp]
\begin{center}
\mbox{
\subfigure[]{
\includegraphics[width=0.4\columnwidth] {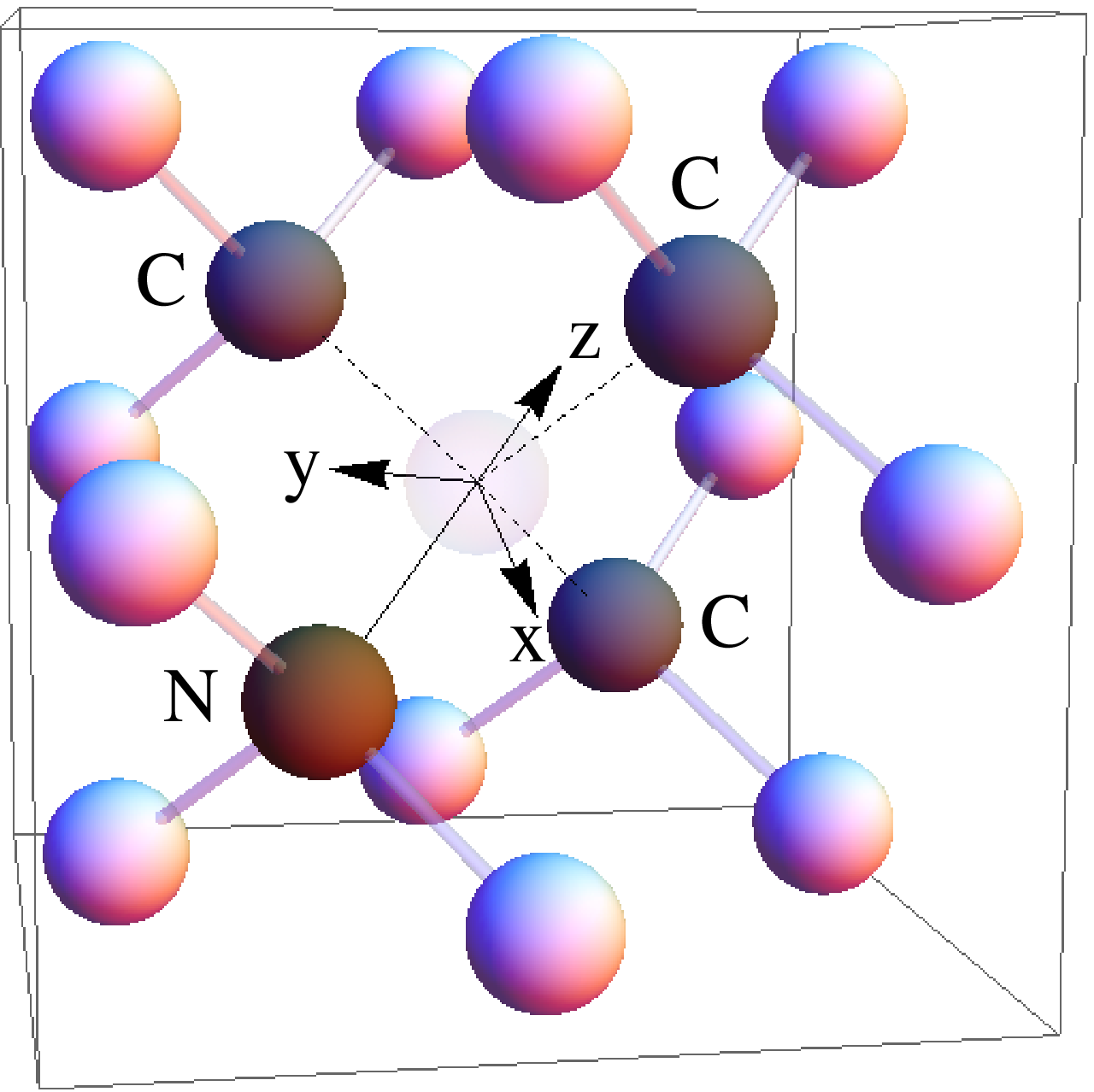}}
\subfigure[]{\includegraphics[width=0.45\columnwidth] {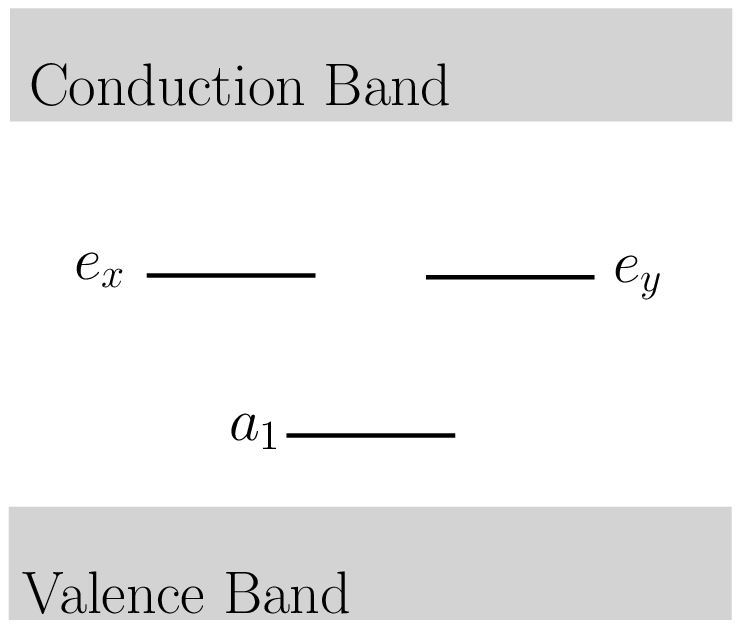}}
}
\mbox{
\subfigure[]{\includegraphics[width=0.75\columnwidth] {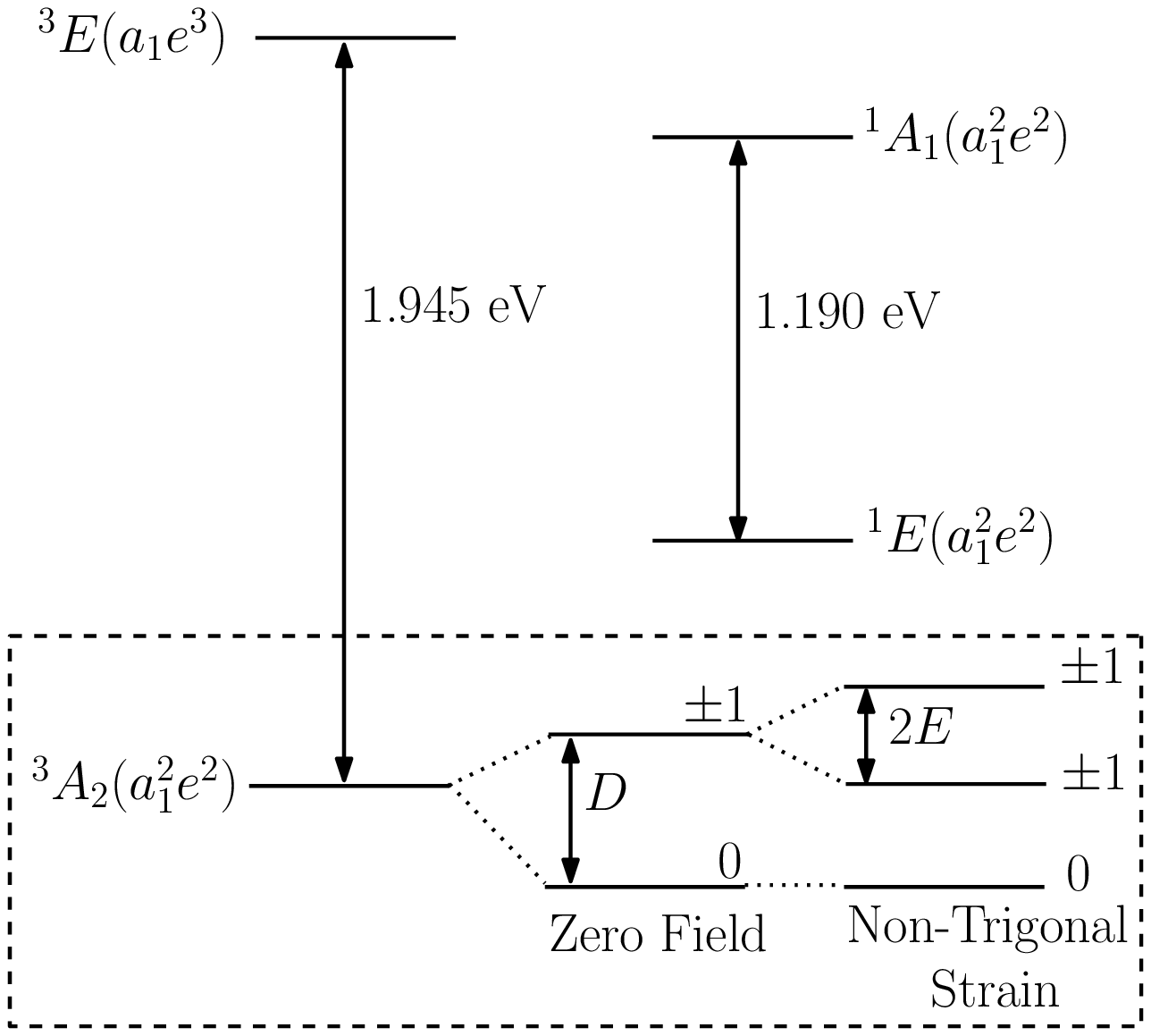}}
}
\caption{(color online) (a) Schematic of the NV center depicting the vacancy (transparent), the nearest neighbor carbon atoms (gray), the substitutional nitrogen atom (brown), the next-to-nearest carbon neighbors (white) and the adopted coordinate system (z axis aligned with the trigonal axis of the center). (b) The NV$^-$ one-electron orbital level structure depicting the diamond valence and conduction bands and the three defect orbitals ($a_1$, $e_x$, and $e_y$) within the bandgap. (c) Schematic of the center's many-electron electronic structure, including the optical 1.945 eV and infrared 1.190 eV ZPL energies. The electronic configurations of the many-electron levels are indicated in parentheses. Inset: The fine structure of the ground $^3A_2$ level: at zero field with a single splitting of $D\sim2.87$ GHz; and in the presence of non-trigonal strain, with a further strain dependent splitting denoted by $E$. }
\label{fig:fig1}
\end{center}
\end{figure}

The NV$^-$ center is a $C_{3v}$ point defect in diamond consisting of a substitutional nitrogen atom adjacent to a carbon vacancy that has trapped an additional electron (refer to figure \ref{fig:fig1}a). As depicted in figure \ref{fig:fig1}b, the one-electron orbital level structure of the NV$^-$ center contains three defect orbital levels ($a_1$, $e_x$ and $e_y$) deep within the diamond bandgap. EPR observations and ab initio calculations indicate that these defect orbitals are highly localized to the center \cite{he93,felton09,larrson08,gali08}. Figure \ref{fig:fig1}c shows the center's many-electron electronic structure generated by the occupation of the three defect orbitals by four electrons \cite{doherty11,maze11}, including the zero phonon line (ZPL) energies of the optical (1.945 eV/637 nm)  \cite{davies76} and infrared (1.190 eV/1042 nm) \cite{rogers08,acosta10b,manson10} transitions. The energy separations of the spin triplet and singlet levels (${^3}A_2\leftrightarrow{^1}E$ and ${^1}A_1\leftrightarrow{^3}E$) are unknown.

As depicted in the inset of figure \ref{fig:fig1}c, the ground $^3A_2$ level exhibits a zero field fine structure splitting between the $m_s=0$ and $\pm1$ spin sub-levels of $D\sim2.87$ GHz (room temperature) due to electron spin-spin interaction \cite{loubser78}. The zero field parameter $D$ and spin quantization axis are thus determined by the trigonal unpaired electron spin density distribution of the ground $^3A_2$ level.  Under crystal strain that distorts the trigonal symmetry of the center, the degeneracy of the $m_s=\pm1$ sub-levels is lifted. The spin-Hamiltonian that describes the $^3A_2$ fine structure is
\begin{equation}
H = D[S_z^2-S(S+1)/3]+E(S_x^2-S_y^2)
\end{equation}
where $E$ is the strain parameter, the $S=1$ spin operators are dimensionless and the $z$ coordinate axis coincides with the center's trigonal symmetry axis (see figure \ref{fig:fig1}a).

The spin of the ground $^3A_2$ level is optically polarized via the combination of optical excitation and spin-selective non-radiative intersystem crossings (ISCs) that preferentially depopulate the $m_s=\pm1$ sub-levels and populate the $m_s=0$ sub-level \cite{manson06}. The ISCs also lead to spin-dependent optical fluorescence that enable the ground state spin to be readout and the performance of ground state optically detected magnetic resonance (ODMR) \cite{manson06}. The dark ISCs of the NV$^-$ center are yet to be fully understood and they are currently believed to be the combined result of spin-orbit coupling of the lowest energy triplet ($^3A_2$, $^3E$) and singlet ($^1A_1$, $^1E$) levels and electron-phonon decay \cite{manson06}.

\begin{figure}[hbtp]
\begin{center}
\mbox{
\subfigure[]{
\includegraphics[width=0.4875\columnwidth] {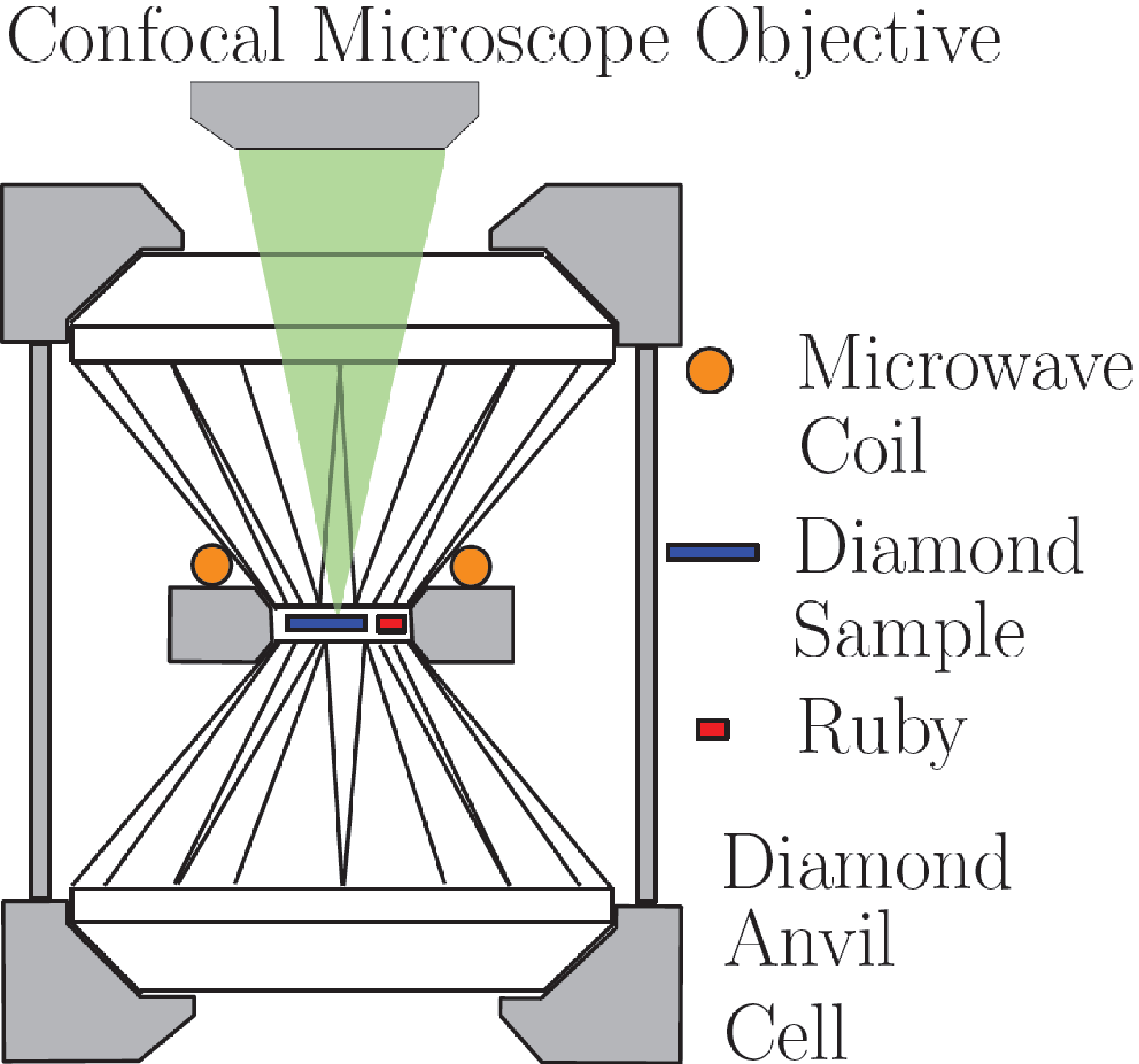}}
\subfigure[]{\includegraphics[width=0.5125\columnwidth] {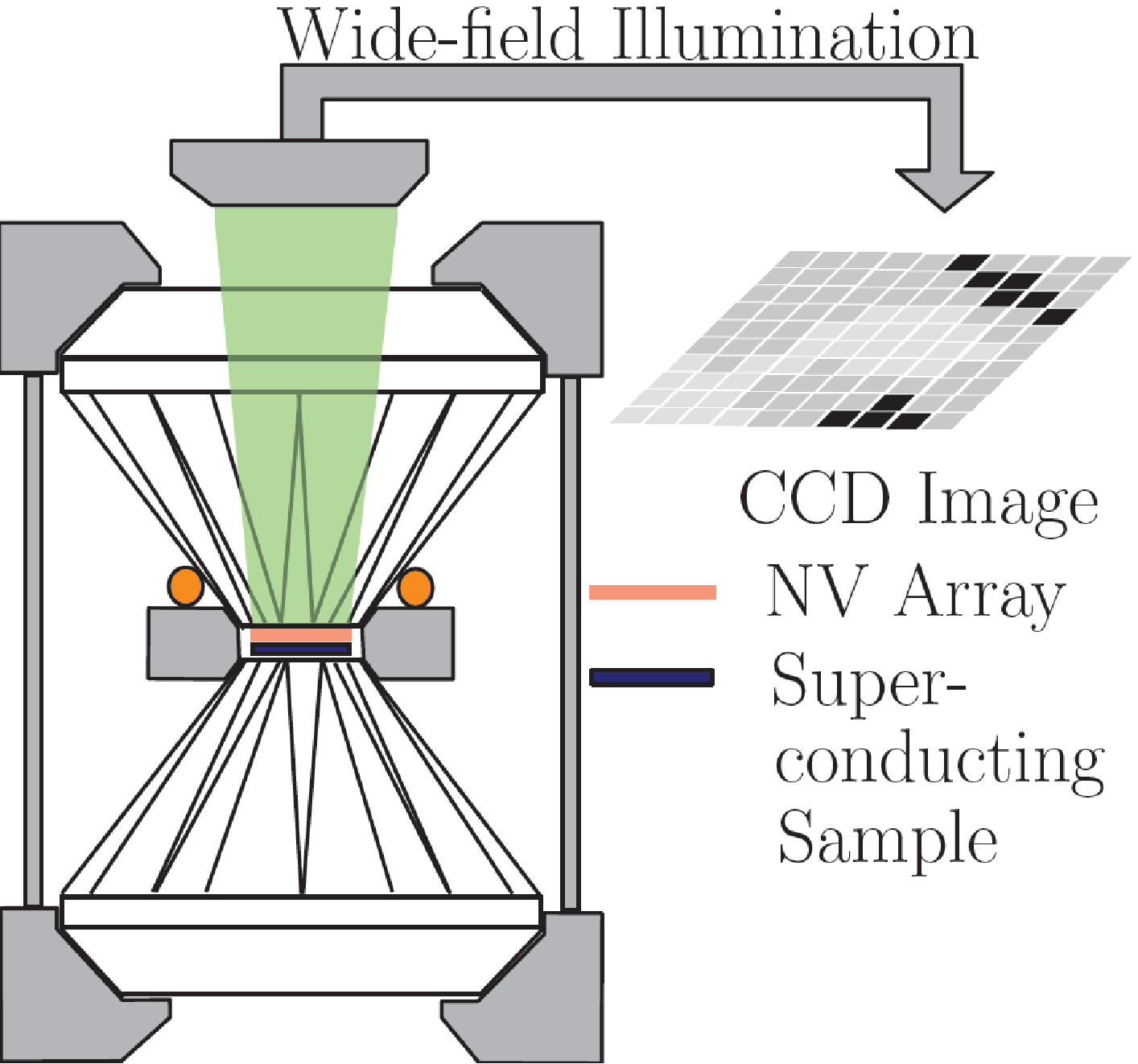}}
}
\caption{(color online) (a) Schematic of the DAC arrangement used in our experiments. Diamond samples were placed in the DAC working space together with the hydrostatic pressure medium and a Ruby for pressure measurement. The diamond culets provide optical access to optically excite (532 nm) and collect the NV$^-$ red-shifted fluorescence. ODMR was performed via microwave excitation provided by a wire or coil embedded in an insulating gasket or wound around a culet. (b) Schematic of the proposed design of a diamond chip containing an NV$^-$ array, which is used to switch between pressure monitoring and wide field imaging of the magnetic field at the surface of a high pressure superconductor.}
\label{fig:DAC}
\end{center}
\end{figure}

In our experiments, we observed the hydrostatic pressure shift of the optical ZPL and the ground state ODMR of small ensembles of NV$^-$ centers at room temperature \cite{supmat}. As depicted in figure \ref{fig:DAC}(a), our experiments were performed in a diamond anvil cell (DAC) using samples of type IIa single-crystal CVD diamonds with low nitrogen content ($<1$ ppm). The shift of the optical ZPL was measured in NaCl pressure media and detected using a spectrometer. We measured the ODMR in two different pressure media: NaCl and Ne.  ODMR was performed by optically exciting the NV$^-$ centers with 532 nm light and scanning the microwave frequency through either a current carrying Pt wire embedded in an insulating BN+epoxy gasket or a copper coil wound a few times around one of the diamond culets. We measured the pressure \textit{in situ} using the fluorescence of a small ruby chip located in the DAC. The pressure calibration from Ref. \onlinecite{mao86} was used. The maximum pressure applied was 60 GPa, which was limited by the blue shift of the NV$^-$ ZPL to the 532 nm optical excitation wavelength. A systematic ODMR temperature shift was avoided in our experiments by allowing the DAC to come to thermal equilibrium in the presence of steady microwave excitation at each pressure.

\begin{figure}[hbtp]
\begin{center}
\mbox{
\subfigure[]{
\includegraphics[width=0.5\columnwidth] {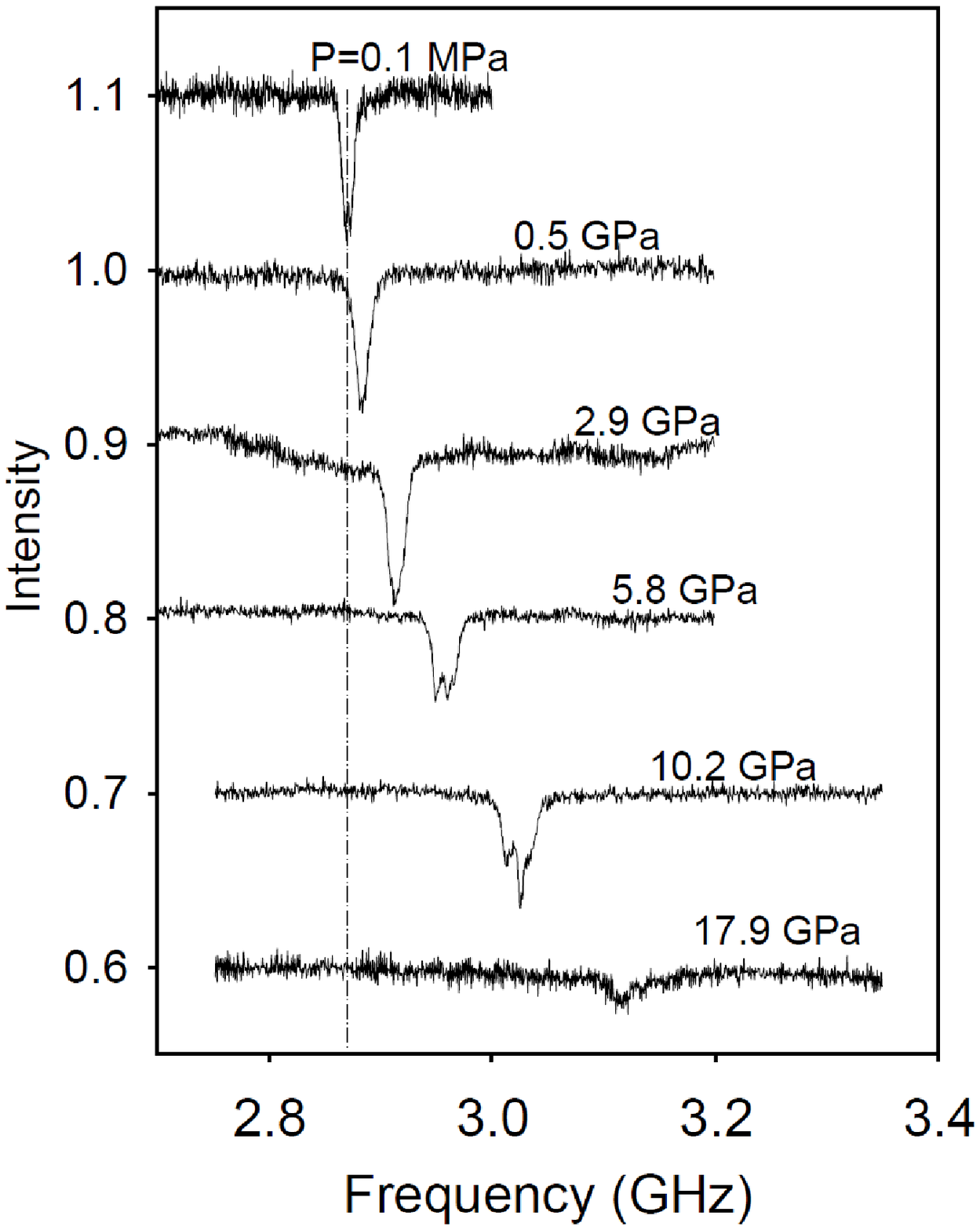}}
\subfigure[]{
\includegraphics[width=0.5\columnwidth] {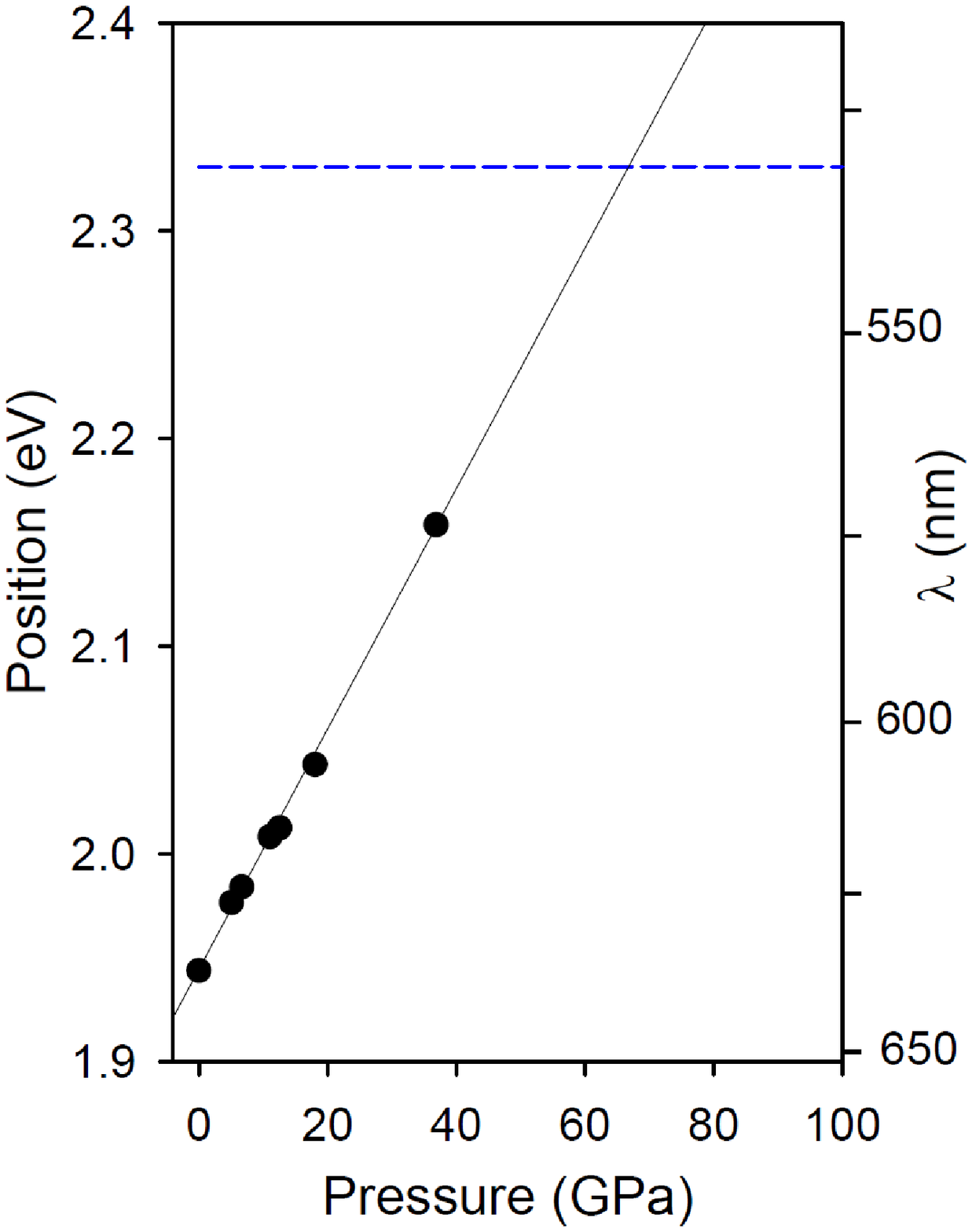}}
}
\mbox{
\subfigure[]{\includegraphics[width=0.75\columnwidth] {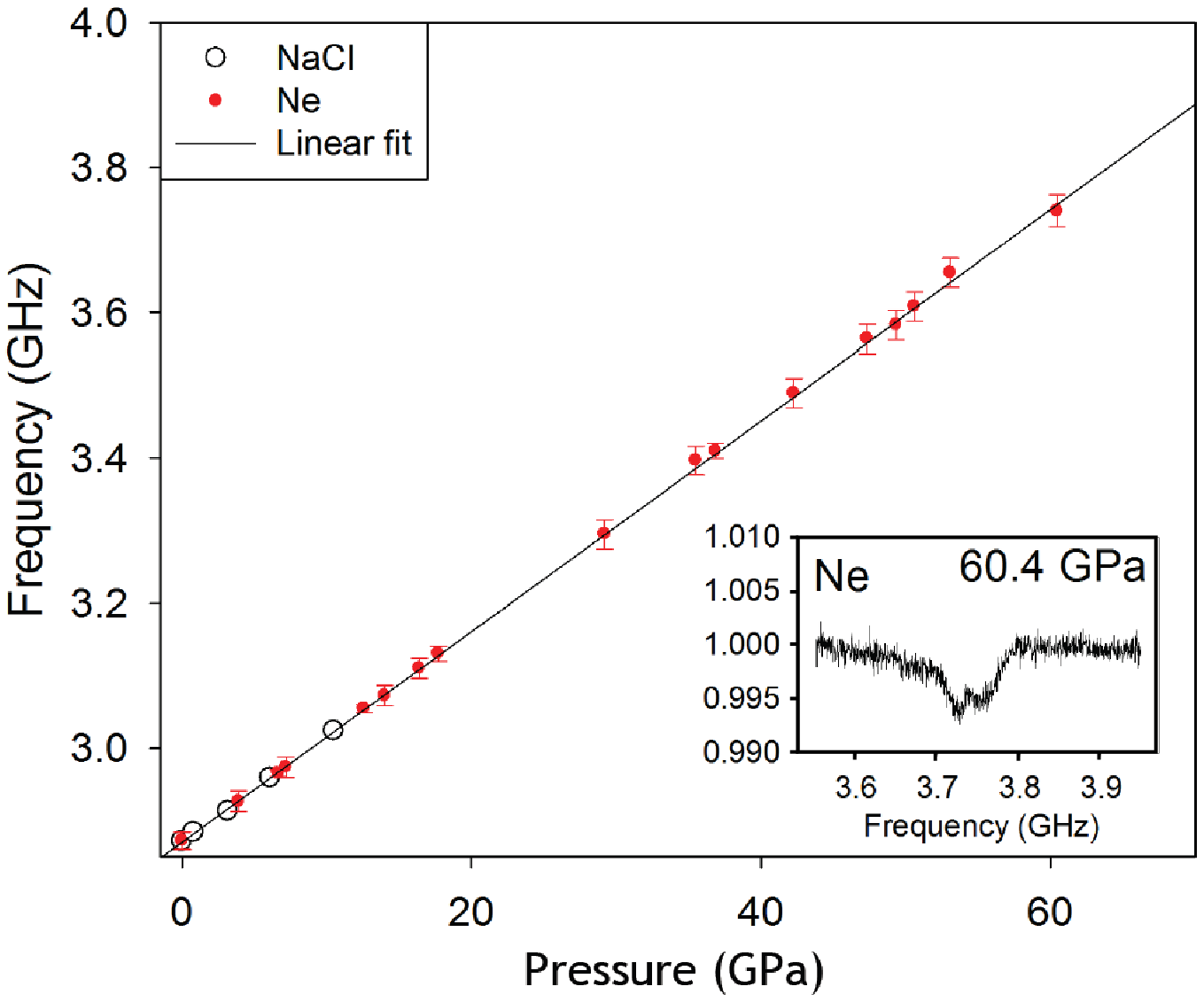}}
}
\caption{(color online) (a) Example NV$^-$ ODMR spectra in the quasihydrostatic NaCl pressure medium. The dashed line indicates the position of the resonance at ambient pressure. (b) The pressure shift of the optical ZPL. Solid line is a linear fit extrapolated to the position of the 532 nm excitation laser. (c) The measured ODMR splitting parameter $D$ in quasihydrostatic (NaCl) and hydrostatic (Ne) pressure media up to 60 GPa. $D$ is taken as the midpoint of the two ODMR peaks and error bars correspond to the full width of the ODMR peaks (see Ref. \onlinecite{supmat} for further details). Inset: the ODMR spectrum in Ne pressure medium at 60 GPa.}
\label{fig:ODMR_results}
\end{center}
\end{figure}

Figure \ref{fig:ODMR_results} depicts selected ODMR spectra and the observed shift of the splitting parameter $D$.   The shift of the $D$ parameter with pressure $P$ is highly linear and has the gradient $d D(P)/ dP =$ 14.58(6) MHz/GPa. The broadening and splitting of the ODMR line at pressures $>4.5$ GPa are results of crystal distortion. At pressures $>4.5$ GPa, the Ne pressure medium freezes at room temperature and becomes quasihydrostatic. As a consequence, the stress applied to the crystal is slightly anisotropic ($<0.4$ GPa at 50 GPa applied pressure) and the crystal distorts \cite{klotz09}. Figure \ref{fig:ODMR_results} also depicts the observed shift of the optical ZPL energy, which is also highly linear and has the gradient 5.75 meV/GPa. This result compares reasonably to a previous measurement of 5.5 meV/GPa obtained from high density ensembles of NV$^-$ centers in type Ib diamond using a N$_2$ pressure medium \cite{kobaryashi93}. The small difference can be attributed to the significantly inhomogeneously strain broadened optical ZPLs of the high density ensembles and the different quasihydrostatic behavior of N$_2$ pressure media.

Our hydrostatic pressure ODMR measurements provided a controlled means to probe the electronic orbitals of the NV$^-$ center. Under hydrostatic pressure, the equilibrium positions of the nuclei of the lattice contract and thus, the defect orbitals change. Considering the ground $^3A_2$ level, the change in the defect orbitals corresponds to the contraction of the unpaired spin density of the level, which thereby increases the electronic spin-spin interactions and the $D$ parameter. Picturing the defect orbitals as linear combinations of atomic orbitals \cite{loubser78}, the contraction of the unpaired spin density maybe seen to have two aspects: (1) the compression of the nuclear lattice decreases the distance between atomic orbitals, and (2) the compression of the nuclear lattice deepens the localizing electrostatic potential of the center, thereby modifying the linear combinations such that the defect orbitals become more localized to the atoms closer to the vacancy. Elaborating on (2), EPR observations and ab initio calculations \cite{he93,felton09,larrson08,gali08} indicate that the defect orbitals are combinations of the orbitals of atoms belonging to concentric shells centered on the vacancy, such that the greatest contributions are from the three nearest neighbor carbon atoms of the vacancy. As pressure is increased, it is pictured that the contributions of orbitals of the inner shell atoms increase as the contributions of more distant shell atoms decrease, thereby resulting in a net shift of electron probability density towards the vacancy.

This picture can be demonstrated using the molecular model of the NV$^-$ center \cite{doherty11,maze11,doherty12}, where the spin-spin $D$ parameter of the $^3A_2$ level is given by
\begin{eqnarray}
D & = & C \bra{e_x(\vec{r}_1)e_y(\vec{r}_2)}\frac{1}{r_{12}^3}-\frac{3z_{12}^2}{r_{12}^5}\left[\ket{e_x(\vec{r}_1)e_y(\vec{r}_2)}\right. \nonumber \\
&& \left.-\ket{e_y(\vec{r}_1)e_x(\vec{r}_2)}\right]
\end{eqnarray}
where $C= 3\mu_0g_e^2\mu_B^2/16\pi h$ is a constant, $\mu_0$ is the permeability of free space, $g_e$ is the free electron g-factor, $\mu_B$ is the Bohr magneton, $h$ is Planck's constant, $\vec{r}_i = x_i\hat{\vec{x}}+y_i\hat{\vec{y}}+z_i\hat{\vec{z}}$ is the position of the $i^{th}$ electron, $r_{12}=|\vec{r}_2-\vec{r}_1|$ and $z_{12}=z_2-z_1$. In the molecular model the defect orbitals of the NV center are approximated by linear combinations of the dangling sp$^3$ atomic orbitals ($n$,$c_1$,$c_2$,$c_3$) of the vacancy's nearest neighbor nitrogen and carbon atoms. The defect orbitals $e_x$ and $e_y$ relevant to the calculation of the $^3A_2$ $D$ parameter take the form
\begin{eqnarray}
e_x & = & \sqrt{\eta} N_x (2c_1-c_2-c_3) + \sqrt{1-\eta}(\ldots) \nonumber \\
e_y & = & \sqrt{\eta} N_y (c_2-c_3) + \sqrt{1-\eta}(\ldots)
\end{eqnarray}
where $\eta$ is the fraction of $e$ orbital probability density associated with the vacancy's nearest neighbor carbon atoms, $N_x=[(2\bra{c_1}-\bra{c_2}-\bra{c_3})(2\ket{c_1}-\ket{c_2}-\ket{c_3})]^{-1/2}$ and $N_y=[(\bra{c_2}-\bra{c_3})(\ket{c_2}-\ket{c_3})]^{-1/2}$ are normalization constants and $(\ldots)$ denote the contributions of atomic orbitals of atoms beyond the vacancy's nearest neighbors. Using the definitions of $e_x$ and $e_y$, considering only terms involving $c_1$, $c_2$, and $c_3$, applying symmetry operations and ignoring orbital overlaps, the $D$ parameter is approximately
\begin{eqnarray}
D & \approx & C\eta^2 \bra{c_1(\vec{r}_1)c_2(\vec{r}_2)}\frac{1}{r_{12}^3}-\frac{3z_{12}^2}{r_{12}^5}\ket{c_1(\vec{r}_1)c_2(\vec{r}_2)}
\end{eqnarray}
This approximation clearly shows that the dominant contribution to the $D$ parameter is a direct integral between the dangling sp$^3$ electronic densities of two carbon atoms surrounding the vacancy. This direct integral is reduced by the unpaired spin density $\eta$ centered on the carbon atoms surrounding the vacancy, which is related to the localization of the $e_x$ and $e_y$ orbitals. Interpretations of EPR studies have estimated $\eta\sim0.84$ \cite{felton09}.

Under hydrostatic pressure, the nuclear lattice is isotropically compressed. The normal nuclear displacement coordinate $Q=P/\gamma B$ of hydrostatic compression is connected to the pressure $P$ via the diamond bulk modulus $B$ and a geometric factor $\gamma$ that relates the hydrostatic volume strain to the nuclear displacement. The linear pressure shift of the $D$ parameter at zero pressure is then
\begin{eqnarray}
\left.\frac{d D}{dP}\right|_0
&\approx&
\frac{C}{\gamma B} \left.\eta^2\right|_0\left.\frac{d}{dQ}\langle\frac{1}{r_{12}^3}-\frac{3z_{12}^2}{r_{12}^5}\rangle\right|_0\nonumber \\
&&+\frac{C}{\gamma B} \left.\frac{d \eta^2}{d Q}\right|_0 \left.\langle\frac{1}{r_{12}^3}-\frac{3z_{12}^2}{r_{12}^5}\rangle\right|_0
\end{eqnarray}
where $\langle\frac{1}{r_{12}^3}-\frac{3z_{12}^2}{r_{12}^5}\rangle=\bra{c_1(\vec{r}_1)c_2(\vec{r}_2)}\frac{1}{r_{12}^3}-\frac{3z_{12}^2}{r_{12}^5}\ket{c_1(\vec{r}_1)c_2(\vec{r}_2)}$. As pictured, the first line accounts for the displacements of the atomic orbitals of the carbon atoms surrounding the vacancy, whereas the second line accounts for the change in the unpaired spin density $\eta$ centred on the carbon atoms.

Due to the approximate nature of the above expression and the difficulty of the direct integrals, we instead considered a semi-classical model of the pressure shift, which yielded the following estimate of the shift due to the displacement of the atomic orbitals (refer to Ref. \onlinecite{supmat} for further details)
\begin{eqnarray}
\frac{C}{\gamma B} \left.\eta^2\right|_0\left.\frac{d}{dQ}\langle\frac{1}{r_{12}^3}-\frac{3z_{12}^2}{r_{12}^5}\rangle\right|_0 & \approx & 6.2 \ \mathrm{MHz/GPa}
\end{eqnarray}
This compares favourably to the observed $\sim15$ MHz/GPa shift. Based upon this estimate, the change in the unpaired spin density maybe inferred
\begin{eqnarray}
\left.\frac{d \eta}{d P}\right|_0= \frac{1}{\gamma B} \left.\frac{d \eta}{d Q}\right|_0\approx 0.0012 \ \mathrm{GPa}^{-1}
\end{eqnarray}
which corresponds to an entirely reasonable $\sim0.1\%$ GPa$^{-1}$ change in the spin density. Clearly, \textit{ab initio} calculations are required to provide a much more accurate model of the spin density variation with pressure. Further experimental observations of the spin density can be achieved via the pressure shifts of the NV$^-$ hyperfine interactions with local $^{13}$C nuclear spins, which directly depend on the spin density at the $^{13}$C nuclei \cite{loubser78}.

Analogous to previous estimates of the field and thermal sensitivities of the NV$^-$ ODMR, the estimated ODMR pressure sensitivity of a single NV$^-$ center at room temperature is \cite{toyli12}
\begin{equation}
\eta_{gs} = \frac{1}{2\pi\frac{d D(P)}{dP} K \sqrt{T_2^\ast}} \sim 0.6 \ \mathrm{MPa/\sqrt{Hz}}
\end{equation}
where $K$ accounts for experimental factors such as collection efficiency. Note that the typical single center room temperature values $K\sim 0.02$ and $T_2^\ast\sim1$ $\mu$s were used in the above estimate. Although the pressure shift of the optical ZPL is much larger than the ODMR, the large homogeneous broadening of the optical ZPL at room temperature and beyond \cite{fu09} ensures that the optical pressure sensitivity is significantly worse than the ODMR sensitivity. However, in the low temperature limit ($T\lesssim 12$ K), where narrow optical lines ($\sim140$ MHz) of engineered NV$^-$ centers can be obtained \cite{stacey12}, the optical sensitivity achieves $\eta_{op}\approx 68$ Pa/$\sqrt{\mathrm{Hz}}$.

We now consider the performance of the current NV$^-$ metrology applications under different conditions. The estimated pressure sensitivity of the NV$^-$ ODMR implies that changes in pressure of $\sim1$ MPa can be detected after one second of averaging time. As the majority of current proposals for NV$^-$ magnetometry and electrometry are targeted at ambient conditions, where pressure fluctuations of the order of $\sim1$ MPa are very unlikely to occur over timescales of a second, the pressure sensitivity of the NV$^-$ ODMR will not influence the performance of these current proposals. Even up to pressures of $\sim50$ GPa, which are relevant to studies of high pressure superconductivity and magnetic phase transitions using DACs \cite{struz02,lin05,takahashi08,xchen10,sun12,kornev05,ahart10}, the pressure fluctuations can be controlled to less than $\sim10$ kPa on timescales of seconds (estimated based on resistance changes at the insulator-metal transition \cite{Ga08}). Hence, the observed hydrostatic pressure shift implies that the current proposals of NV$^-$ magnetometry and electrometry will retain much of their sensitivity in the extremes of pressure.

Furthermore, the NV$^-$ ODMR offers a more sensitive means to measure high pressure than existing techniques. As employed in this work, the current technique of choice utilizes the optical transitions of ruby and has a typical accuracy of $\sim10$ MPa \cite{klotz09}. The NV$^-$ center's pressure sensitivity, its possible inclusion with the sample in the DAC and its capability to switch to magnetometry or electrometry modes (assuming pressure fluctuations are controlled to $<1$ MPa on the timescale of seconds) undoubtedly make the NV$^-$ center the ideal probe for high pressure phenomena. One possible design is depicted in figure \ref{fig:DAC}(b), where a thin diamond chip containing an array or ensemble of NV$^-$  centers is placed on top of a high pressure superconductor sample inside the DAC. Employing existing techniques of wide field NV$^-$ magnetometry \cite{steinert10}, the NV$^-$ array can be switched between pressure and stress \cite{Graz11} sensing and magnetometry, offering an unprecedented means to monitor the pressure and image the magnetic phenomena occurring at the surface of the high pressure superconductor.

\begin{acknowledgements}
High-pressure experiments were supported by BES/DOE under grant number DE-FG02-02ER45955. CVD diamond preparation was supported by DOE-NNSA under grant number DE-FC52-08NA28554. This work was also supported by the Australian Research Council Discovery Project (DP120102232)and Centre of
Excellence for Quantum Computation and Communication Technology (project number CE110001027). L.P. McGuinness wishes to acknowledge support from the Alexander von Humboldt foundation.
\end{acknowledgements}

\end{document}